\documentclass[preprint,aps,showpacs]{revtex4}
\usepackage{mathrsfs}
\usepackage{amsmath}
\usepackage{amssymb}
\usepackage{epsfig}
\usepackage{graphicx}
\usepackage{booktabs}
\usepackage{subfigure}
\usepackage{color}

\begin{document}

\title{Revisiting the $B \to \pi \rho$, $\pi\omega $ Decays in the
Perturbative QCD Approach Beyond the Leading Order}
\author{Zhou Rui}
\author{Gao Xiangdong}
\author{Cai-Dian L\"{u}}\email{lucd@ihep.ac.cn}
\affiliation{Institute  of  High  Energy  Physics  and  Theoretical  Physics Center for Science Facilities,
Chinese Academy of Sciences, Beijing 100049, People's Republic of China }

\date{\today}

\begin{abstract}
We calculate the branching ratios and CP asymmetries of the $B \to
\pi \rho$, $\pi\omega $ decays in the perturbative QCD factorization
approach up to the next-to-leading-order contributions. We find that
the next-to-leading-order contributions can interfere with the
leading-order part constructively or destructively for different
decay modes. Our numerical results have a much  better agreement
with current available data than previous leading-order
calculations, e.g., the next-to-leading-order corrections enhance
the $B^0\rightarrow \pi^0\rho^0$ branching ratios by a factor 2.5,
which is helpful to narrow the gaps between theoretic predictions
and experimental data. We also update the direct CP-violation
parameters, the mixing-induced CP-violation parameters of these modes,
which show a better agreement with experimental data than many of the
other approaches.

\end{abstract}


\maketitle

\section{Introduction}

The charmless B meson decays are not only suitable to study CP
violations but also sensitive to new physics\cite{iiba}. During the
past decade, the B factory experiments achieved great
successes. Furthermore, the current LHC experiments will provide 2--3 orders more
B meson events than the B factories \cite{lhcb1}. A large
number of rare $B$ meson decay channels will be measured by the
future super B factories. The research on the charmless decays of
$B$ meson is therefore becoming more interesting than ever before
\cite{lhcb2}.

The theoretical calculations of color-suppressed decay channels,
such as $B^0 \to \pi^0 \pi^0$, met a difficulty for a relatively
much smaller branching ratios than the experimental measurements
\cite{prl831914,prd63074009,pdg2010}. The difference between direct
CP-asymmetry measurement of $B^0 \to K^+\pi^-$ and $B^+\to K^+\pi^0$
showed a very large discrepancy between the leading-order (LO)
theoretical calculations and experimental data, which induced a lot
of new physics discussions \cite{pikpuzzle}. One of the standard
model solutions to this puzzle also requires large color-suppressed
tree amplitudes \cite{prd114005}. Some of the next-to-leading-order (NLO)
QCD calculations in the perturbative QCD factorization approach
(pQCD) \cite{prd114005,prd094020,0807,prd114001} show that the NLO
contributions can significantly change the LO predictions for some
decay modes, especially the color-suppressed modes. It is therefore
necessary to calculate the NLO corrections to those two-body
charmless B meson decays in order to improve the reliability of the
theoretical predictions.

The $B \to \pi \rho$ decays, which are helpful for the determination
of the Cabibbo--Kobayashi--Maskawa(CKM) unitary triangle $\alpha$
angle measurement in addition to the $B\to \pi\pi$ decays, have a
much more complication. Either of $B^0$ or $\bar B^0$ meson can
decay to both the $\pi^-\rho^ +$ and $\pi^+\rho^-$ final states,
which lead to altogether four decay amplitudes. Since $B^0$ and
$\bar B^0$ meson mix easily, these channels exhibit unique features
of mixing and decay interference in B physics. The recent B factory
measurements indeed show that the interesting phenomenology with
possible large direct CP asymmetry \cite{exp}. Unlike the branching
ratios, the CP asymmetries are sensitive to high order
contributions. Similar to the color-suppressed $B^0 \to \pi^0\pi^0$
mode, the neutral decay modes $B^0 \to \pi^0 \rho^0$, $\pi^0\omega$
are also expected to receive considerable NLO contributions.
Therefore, it is necessary to calculate NLO corrections to the $B
\to \pi \rho, \pi\omega $ decays in the pQCD approach for the reason
that previous pQCD calculations \cite{epjc23275} are already too old
with only LO accuracy. In this paper, we calculate the NLO
contributions arising from the vertex corrections, the quark loops
and the chromo-magnetic penguin operator $O_{8g}$. Combining our
results with the NLO accuracy Wilson coefficients and Sudakov
suppression factors, we present a numerical analysis of $B \to \pi
\rho$, $\pi\omega$ decays.

Our paper is organized as follows: we first review the pQCD
factorization approach in Sec.~\ref{sec:f-work}. Then, in
Sec.~\ref{sec:nlo}, we show our analytical results of NLO
calculations. The numerical results are given in
Sec.~\ref{sec:numerical}. Finally we close this paper with a
conclusion.

\section{ Theoretical framework}\label{sec:f-work}

For the studied $B \to \pi \rho,\pi\omega $ decays, the weak
effective Hamiltonian $\mathcal {H}_{eff}$ for $b\rightarrow d$ transition can
be written as
\begin{eqnarray}\label{eq:heff}
\mathcal {H}_{eff}=\frac{G_F}{\sqrt{2}}[\xi_u (C_1(\mu)O_1^u(\mu)
+C_2(\mu)O_2^u(\mu))-\xi_t \sum_{i=3}^{10}C_i(\mu)O_i(\mu)]
\end{eqnarray}
where $\xi_u =V_{ub}V^*_{ud}$, $\xi_t = V_{tb}V^*_{td}$ are the CKM
matrix elements. $O_i(\mu)$ and $C_i(\mu)$ are the four-quark
operators and corresponding Wilson coefficients, respectively.
Expressions of $C_i$ and $O_i$ can be found in Ref.\cite{rmp681125}.
In the following, we will use this effective Hamiltonian to calculate
decay amplitudes in the pQCD approach. So, we first give a brief
review of pQCD approach and present relevant wave functions.

\subsection{pQCD factorization approach}

In the framework of the pQCD factorization, three scales are
involved in the non-leptonic decays of B mesons: the weak
interaction scale $m_W$, the hard subprocess scale $t$ and the
transverse momenta of the constituent quark $k_T$. The large logs
between W boson mass scale and the hard scale $t$ have been resummed
by the renormalization group equation method to give the effective
Hamiltonian of four-quark operators. In two-body charmless hadronic
B decays, the final state meson masses are negligible compared with
the large B meson mass. Therefore the constituent quarks in the
final state mesons are collinear objects in the rest frame of B
meson. The momentum of light quark in B meson is at the order of
$\Lambda_{QCD}$, such that a hard gluon is needed to transfer energy
to make it a collinear quark into the final state meson. These
perturbative calculations meet end-point singularity in dealing with
the meson distribution amplitudes at the end-point. Usually in the
collinear factorization approaches such as QCD
factorization\cite{prl831914} and soft-collinear effective
theory\cite{prd63114020}, people parameterize this kind of decay
amplitudes into free parameters to fit the data. While in the
perturbative QCD factorization approach, we take back the parton
transverse momentum $k_T$ to regulate this divergence.

In the pQCD approach, the decay amplitude $A(B \rightarrow M_2M_3)$
can be written conceptually as the convolution \cite{prd69094018}
\begin{eqnarray}\label{eq:AAAA}
\mathcal {A}(B \rightarrow M_2M_3)=\int d^4k_1d^4k_2d^4k_3
\text{Tr}[C(t)\Phi_B(k_1)\Phi_{M_2}(k_2)\Phi_{M_3}(k_3)H(k_1,k_2,k_3,t)]
\end{eqnarray}
where $k_i$ are momenta of light quarks included in each meson, and
Tr denotes the trace over Dirac and color indices. The hard function
$H(k_1, k_2, k_3, t)$ describes the four-quark operator and the
spectator quark connected by a hard gluon of order  $\bar {\Lambda}
M_B$, which can be calculated perturbatively. The energy scale $t$
is chosen as the maximal virtuality of internal particles in a hard
amplitude, in order to suppress higher order
corrections\cite{prd074004}. $\Phi_{M_i}$ is the wave function of
meson $M_i$. The hard kernel $H$ depends on the processes
considered, while the wave functions $\Phi_{M_i}$ are process
independent that can be extracted from other well measured
processes, so one can make quantitative predictions here.

It is convenient to work at the B meson rest frame and the light
cone coordinate. The final state meson $M_2$ is moving along the
direction of $v = (0, 1, \bf 0_T)$ and $M_3$ is along $n = (1, 0,
\bf 0_T)$. Here we use $x_i$ to denote the momentum fractions of
anti-quarks in mesons, and $\bf k_{iT}$ to denote the transverse
momenta of the anti-quarks. The mass of light meson ($\pi$) is
neglected. After integration over $k_1^-$, $k_2^-$ and $k_3^+$ in
Eq.(\ref{eq:AAAA}), we are led to
\begin{eqnarray}\label{eq:factorization}
\mathcal {A}&=&\int dx_1dx_2dx_3b_1db_1b_2db_2b_3db_3\nonumber\\&&
\text{Tr}[C(t)\Phi_B(x_1,b_1)\Phi_{M_2}(x_2,b_2)\Phi_{M_3}(x_3,b_3)H(x_i,b_i,t)S_t(x_i)\exp(-S(t))]
\end{eqnarray}
where $b_i$ are the conjugate variables of $\bf k_{iT}$. The jet
function $S_t(x_i)$ arises from the threshold resummation of the
large double logarithms $\ln^2(x_i)$, and the Sudakov exponent
$S(t)$ comes from the double logarithms of collinear and soft
divergences.

\subsection{Wave Functions}

There are generally two Lorentz structures in the B
meson distribution amplitudes, which can be decomposed as \cite{qiaocf}
\begin{eqnarray}\label{eq:bbwave1}
\int_0^1\frac{d^4z}{(2\pi)^4}e^{ik_1\cdot z}\langle 0|
\bar{b}_{\alpha}(0)d_{\beta}(z)|B(p_B)\rangle=-\frac{i}{\sqrt{2N_c}}
[(\rlap{/}{p_B}+m_B)\gamma_5(\phi_B(k_1)-\frac{\rlap{/}{n}-\rlap{/}{v}}{\sqrt{2}}\bar{\phi}_B(k_1))].
\end{eqnarray}
With $N_c=3$,
they obey the following normalization conditions:
\begin{eqnarray}
\int\frac{d^4k_1}{(2\pi)^4}\phi_B(k_1)=\frac{f_B}{2\sqrt{2N_c}},\quad \int\frac{d^4k_1}{(2\pi)^4}\bar{\phi}_B(k_1)=0.
\end{eqnarray}
However, the contribution of $\bar{\phi}_B$ is numerically neglected\cite{epjc28515}.
 Therefore, we will only consider the contributions from $\phi_B$. In b
space the B meson wave function can be expressed by
\begin{eqnarray}\label{eq:bwave1}
\Phi_B(x,b)=\frac{1}{\sqrt{2N_c}}(\rlap{/}{P_B}+m_B)\gamma_5\phi_B(x,b).
\end{eqnarray}

For the light pseudo-scalar mesons $\pi$, the wave function can be
defined as \cite{zpc48239}
\begin{eqnarray}\label{eq:piwave}
\Phi(P,x,\xi)=\frac{i}{\sqrt{2N_c}}\gamma_5[\rlap{/}{P}\phi^A(x)+m_0\phi^P(x)+\xi
m_0(\rlap{/}{n}\rlap{/}{v}-1)\phi^T(x)],
\end{eqnarray}
where $ P$ is the momentum of the light meson $\pi$, $m_0$ is the chiral mass which is defined using the meson mass
$m_P$ and the quark masses as $m_0=m^2_P/(m_{q_1}+m_{q_2})$. $x$ is the
momentum fraction of the quark (or anti-quark) inside the meson,
respectively. When the momentum fraction of the quark (anti-quark)
is set to be $x$, the parameter $\xi$ should be chosen as $+1$$ (-1)$.

For the considered decays, the vector meson $V(\rho,\omega)$ is
longitudinally polarized. The longitudinal polarized component of
the wave function is defined as:
\begin{eqnarray}\label{eq:vwave}
 \Phi_V=\frac{1}{\sqrt{2N_c}}[\rlap{/}{\epsilon}(m_V\phi_V(x)+\rlap{/}{P_V}\phi_V^t(x))+m_V\phi_V^s(x)],
\end{eqnarray}
where the polarization vector $\epsilon$ satisfies $P_V\cdot
\epsilon=0$.

\section{Analytical calculations}\label{sec:nlo}

\begin{figure}[!htbh]
\begin{center}
\vspace{-2cm} \centerline{\epsfxsize=12 cm \epsffile{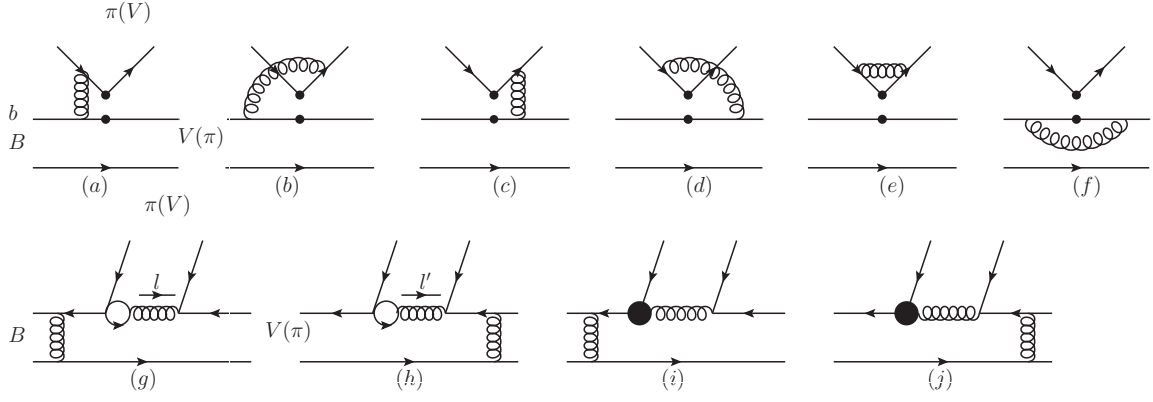}}
\vspace{-8cm} \caption{NLO corrections to the  hard kernels. The
diagrams (a--f),  (g,h) and   (i,j) are commonly called vertex
corrections, quark-loop corrections, and chromo-magnetic penguin
corrections,   respectively.}. \label{fig:nlodiagram}
\end{center}
\end{figure}

 Our NLO corrections for pQCD approach
include the following parts:
\begin{itemize}
\item The NLO hard kernel $H^{(1)}(x_i,b_i,t)$, which  includes the vertex corrections, the quark loops and chromo-magnetic penguins.
\item The NLO Wilson coefficients $C^{NLO}(t)$, which have been calculated in the literature \cite{rmp681125}.
\item  The exponential Sudakov factor $\exp[-S^{NLO}(t)]$ includes the  Sudakov factor $s(P,b)$ and
renormalization group running factor $g_2(t,b)$.
\end{itemize}
So, at the NLO, Eq.(\ref{eq:factorization}) can be written as
\begin{eqnarray}\label{eq:amnlo}
\mathcal {A}^{NLO}&=&\int dx_1dx_2dx_3b_1db_1b_2db_2b_3db_3
\text{Tr}[C^{NLO}(t)\Phi_B(x_1,b_1)\Phi_{M_2}(x_2,b_2)\Phi_{M_3}(x_3,b_3)\nonumber\\&&
(H^{(0)}(x_i,b_i,t)+H^{(1)}(x_i,b_i,t))S_t(x_i)\exp(-S^{NLO}(t))].
\end{eqnarray}

We will give these calculations in the following of this section.

\subsection{Vertex corrections}

The vertex corrections are part of the complete  NLO   Wilson
coefficients for four-quark operators, which cancel the explicit
renormalization scale $\mu$ dependence of the Wilson coefficients.
The vertex correction diagrams are illustrated by
Figs.\ref{fig:nlodiagram}(a)--\ref{fig:nlodiagram}(f), among which
Fig.(e) and (f) are new compared to  the QCDF calculation \cite{prl831914}.
Here, we have introduced transverse momentum $k_T$ in regularizing
the infrared divergence. Our results are different from the QCDF
approach\cite{prl831914} for different regularization schemes.

 The vertex corrections
 to the $B\rightarrow \pi \rho,\pi \omega$ decays modify the Wilson coefficients
 for the emission amplitudes  into
\begin{eqnarray}\label{eq:vertex}
a_1(\mu)& \rightarrow  & a_1(\mu)+\frac{\alpha_s(\mu)}{4\pi}C_F[\frac{C_1(\mu)}{N_c}V_1(M)+C_2(\mu)V'_1(M)], \nonumber\\
a_2(\mu)& \rightarrow  & a_2(\mu)+\frac{\alpha_s(\mu)}{4\pi}C_F[\frac{C_2(\mu)}{N_c}V_2(M)+C_1(\mu)V'_2(M)], \nonumber\\
a_i(\mu)& \rightarrow  & a_i(\mu)+\frac{\alpha_s(\mu)}{4\pi}C_F[\frac{C_{i\pm 1}(\mu)}{N_c}V_i(M)+C_i(\mu)V'_i(M)],\quad i=3-10,
\end{eqnarray}
where $M$ denotes the meson emitted from the weak vertex, and the
upper (lower) sign applies for odd (even) $i$. When the emitted
meson $M$ is a pseudo-scalar meson, the functions $V_i(M)$ and
$V'_i(M)$ are given by
\begin{eqnarray}\label{eq:vm}
V_i(M)&=&
\left\{ \begin{array}{lll}
8 \ln \frac{m_b}{\mu} -18 +\frac{\pi^2}{3}-6i\pi
+\frac{2\sqrt{2N_c}}{f_M}\int_0^1dx\phi_M^A(x)g_1(x), & \quad for \quad $i=1-4,9,10$ \\
 -16 \ln \frac{m_b}{\mu} +6 +\frac{\pi^2}{3}
+\frac{2\sqrt{2N_c}}{f_M}\int_0^1dx\phi_M^A(x)g_2(x), & \quad for \quad  $i=5,7$\\
-16 \ln \frac{m_b}{\mu} +6 +\frac{\pi^2}{3}
+\frac{2\sqrt{2N_c}}{f_M}\int_0^1dx\phi_M^P(x)h(x) & \quad for \quad $i=6,8$
 \end{array} \right.\nonumber\\
V'_i(M)&=&
\left\{ \begin{array}{lll}
-4 \ln \frac{m_b}{\mu}+\frac{\pi^2}{3}-3i\pi, & \quad for \quad $i=1-4,5,7,9,10$ \\
 -16 \ln \frac{m_b}{\mu}+12+\frac{\pi^2}{3} & \quad for \quad $i=6,8$
 \end{array} \right.
\end{eqnarray}
where $m_b$ is the mass of b quark. The functions $g_1(x)$, $g_2(x)$ and $h(x)$ are given as
\begin{eqnarray}\label{eq:vm1}
g_1(x)&=&3\ln x+3\ln (1-x)+\frac{\ln(1-x)}{x}-2\frac{\ln x}{1-x}\nonumber\\&&
+[\ln^2 x+2 \text{Li}_2(\frac{x}{x-1})+4i\pi \ln x-(x \rightarrow 1-x)],\nonumber\\
g_2(x)&=&-3\ln x-3\ln (1-x)+2\frac{\ln(1-x)}{x}+\frac{\ln x}{1-x}\nonumber\\&&
+[\ln^2 x+2 \text{Li}_2(\frac{x}{x-1})+4i\pi \ln x-(x \rightarrow 1-x)],\nonumber\\
h(x)&=&
\ln^2 x+2 \text{Li}_2(\frac{x}{x-1})+\frac{\ln x}{2(1-x)}+4i\pi \ln x-(x \rightarrow 1-x).
\end{eqnarray}

When a vector meson $V(V=\rho,\omega)$ is emitted from the weak
vertex, $\phi_M^A(\phi_M^P)$ is replaced by $\phi_V(-\phi_V^s)$, and
$f_M$ by $f_V^T$ in the third line of Eq.(\ref{eq:vm}). Note that,
the amplitude $ F^P_{e\pi} $ from the operators $O_{5-8}$ vanishes
at LO, because neither the scalar nor the pseudo-scalar density
gives contributions to the vector meson production, i.e.
$<V|S+P|0>=0$. On including the vertex corrections, the NLO piece
$a_{VC}$, containing the vertex-correction  of $a_{6,8}$ in
Eq.(\ref{eq:vertex}), contributes through the following additional
amplitudes\cite{prd094020}:
\begin{eqnarray}\label{eq:addition}
 f_V F^P_{e\pi}\rightarrow a_{VC}f^T_V F^P_{e\pi}+f_V F_{e\pi}
\end{eqnarray}
where $F_{e\pi}$ is the decay amplitude of factorizable emission
diagrams with the structure of $(V-A) \otimes (V-A)$ insertion;
while $F^P_{e\pi}$ is the corresponding decay amplitude  with
$(S-P) \otimes (S+P)$insertion.

\subsection{Quark loops}

The contributions from the quark loops are illustrated by
Fig.\ref{fig:nlodiagram}(g)-\ref{fig:nlodiagram}(h). The quark-loop
contributions are generally called the Bander--Silver--Soni mechanism
\cite{prl43242}, which plays a very important role in producing the
direct CP-violation strong phase in the QCDF/SCET approaches. We
include quark-loop amplitudes from the up-, charm-, and
QCD-penguin-loop corrections, the quark loops from the electroweak
penguins are neglected due to their smallness.

For the $b\rightarrow d$ transition, the contributions from the various quark loops are described
by the effective Hamiltonian $\mathcal
{H}^{(ql)}_{eff}$\cite{prd114005},
\begin{eqnarray}\label{eq:qlheff}
\mathcal {H}^{(ql)}_{eff}&=&-\sum_{q=u,c}\sum_{q'}\frac{G_F}{\sqrt{2}}V_{qb}V^*_{qd}\frac{\alpha_s(\mu)}{2\pi}C^{(q)}(\mu,l^2)
(\bar{d}\gamma_{\rho}(1-\gamma_5)T^ab)(\bar{q'}\gamma^{\rho}T^aq') \nonumber\\&&
+\sum_{q'}\frac{G_F}{\sqrt{2}}V_{tb}V^*_{td}\frac{\alpha_s(\mu)}{2\pi}C^{(t)}(\mu,l^2)
(\bar{d}\gamma_{\rho}(1-\gamma_5)T^ab)(\bar{q'}\gamma^{\rho}T^aq'),
\end{eqnarray}
with
\begin{eqnarray}\label{eq:cq}
C^{(q)}(\mu,l^2)&=&[G^{(q)}(\mu,l^2)-\frac{2}{3}]C_2(\mu),\nonumber\\
C^{(t)}(\mu,l^2)&=&[G^{(s)}(\mu,l^2)-\frac{2}{3}]C_3(\mu)+\sum_{q''=u,d,s,c}G^{(q'')}(\mu,l^2)[C_4(\mu)+C_6(\mu)],
\end{eqnarray}
where $l^2$ being the invariant mass of the intermediate gluon,
which connects the quark loops with the $\bar{q'}q$  pair. 
Because of the absence of the end-point singularities associated
with $l^2,l'^2\rightarrow 0$, we have dropped the parton transverse
momenta $k_T$ in $l^2,l'^2$ for simplicity.
 The integration function $G^{(q)}(\mu,l^2)$
for the loop of the quarks $q = (u, d, s, c)$ is defined as
\begin{eqnarray}\label{eq:gq}
G^{(q)}(\mu,l^2)=-4\int_0^1dx
x(1-x)\ln\frac{m_q^2-x(1-x)l^2}{\mu^2}.
\end{eqnarray}

Finally, the quark-loop contributions shown in
Fig.\ref{fig:nlodiagram}(g) and \ref{fig:nlodiagram}(h) to the
considered $B\rightarrow \pi V$ decays with $V=\rho , \omega$ can be
written as
\begin{eqnarray}\label{eq:mql}
\mathcal {A}^{(ql)}_{V\pi}=\langle V\pi|\mathcal
{H}^{(ql)}_{eff}|\bar B\rangle= \sum_{q=u,c,t}\xi^*_q[\mathcal
{M}^{(q)}_{V\pi}+\mathcal {M}^{(q)}_{\pi V}].
\end{eqnarray}
 The two kinds of topological decay amplitude of the $ B \rightarrow V$ or $ B \rightarrow \pi$  transition are written as
\begin{eqnarray}\label{eq:qlpi}
\mathcal {M}^{ql}_{V\pi
}&=&\frac{4}{\sqrt{3}}C_F^2m_B^4\int_0^1dx_1dx_2dx_3\int_0^{\infty}
b_1db_1b_2db_2\phi_B(x_1,b_1) \nonumber\\&& \times
\left\{[-(1+x_2)\phi_{V}(x_2)\phi^A_{\pi}(x_3)+2r_{\pi}\phi_{V}(x_2)\phi^P_{\pi}(x_3)\right.
\nonumber\\&& \left.
-(1-2x_2)r_V\phi^A_{\pi}(x_3)(\phi^s_{V}(x_2)+\phi^t_{V}(x_2))
\right. \nonumber\\&& \left.
+2(2+x_2)r_{\pi}r_{V}\phi^s_{V}(x_2)\phi^P_{\pi}(x_3)
-2x_2r_{\pi}r_{V}\phi^t_{V}(x_2)\phi^P_{\pi}(x_3) ]\right.
\nonumber\\&& \left. \times
\alpha_s^2(t_1)h_{ql}(x_1,x_2,b_1,b_2)C^{(q)}(t_1,l^2)\exp[-S_{ql}(t_1)]\right.
\nonumber\\&& \left.
+2r_V(2r_{\pi}\phi^P_{\pi}(x_3)-\phi^A_{\pi}(x_3))\phi^s_{V}(x_2)\right.
\nonumber\\&& \left. \times
\alpha_s^2(t_2)h_{ql}(x_2,x_1,b_2,b_1)C^{(q)}(t_2,l^2)\exp[-S_{ql}(t_2)]
\right\},
\end{eqnarray}
\begin{eqnarray}\label{eq:qlv}
\mathcal {M}^{ql}_{\pi
V}&=&\frac{4}{\sqrt{3}}C_F^2m_B^4\int_0^1dx_1dx_2dx_3\int_0^{\infty}
b_1db_1b_2db_2\phi_B(x_1,b_1)\nonumber\\&&
\left\{[(1+x_2)\phi^A_{\pi}(x_2)\phi_{V}(x_3)
+(1-2x_2)r_{\pi}(\phi^P_{\pi}(x_2)+\phi^T_{\pi}(x_2))\phi_{V}(x_3)\right.
\nonumber\\&& \left.
2r_{V}r_{\pi}(2+x_2)\phi_{\pi}^P(x_2)\phi_{V}^s(x_3)-
2r_{V}x_2\phi_{\pi}^T(x_2)\phi_{V}^s(x_3)\right. \nonumber\\&&
\left. +2r_{V}\phi_{\pi}^A(x_2)\phi_{V}^s(x_3)] \times
\alpha_s^2(t_1)h_{ql}(x_1,x_2,b_1,b_2)C^{(q)}(t_1,l^2)\exp[-S_{ql}(t_1)]\right.
\nonumber\\&& \left.
+2r_{\pi}[\phi^P_{\pi}(x_2)\phi^s_{V}(x_3)+2\phi^P_{\pi}(x_2)\phi^s_{V}(x_3)]\right.
\nonumber\\&& \left. \times
\alpha_s^2(t_2)h_{ql}(x_2,x_1,b_2,b_1)C^{(q)}(t_2,l^2)\exp[-S_{ql}(t_2)]
\right\},
\end{eqnarray}
where the ratios $r_V=m_V/m_B, r_{\pi}=m_0^{\pi}/m_B$. The hard scales and the gluon invariant masses are given by
\begin{eqnarray}\label{eq:t1t2}
t_1 &=& \max (\sqrt{x_2}m_B, \sqrt{x_1 x_2}m_B, \sqrt{x_3(1-x_2)}m_B,1/b_1,1/b_2),\nonumber\\
t_2 &=& \max (\sqrt{x_1}m_B, \sqrt{x_1 x_2}m_B, \sqrt{|x_3-x_1|}m_B,1/b_1,1/b_2),\nonumber\\
l^2&=& x_3(1-x_2)m_B^2,\quad l'^2=(x_3-x_1)m_B^2.
\end{eqnarray}
The hard functions $h_{ql}$ are included in the appendix.

\subsection{Chromo-magnetic penguins}

The chromo-magnetic penguin contributions are of NLO in $\alpha_s$
within the pQCD formalism.  They are at the same order in $\alpha_s$
as the penguin contributions.

According to ref.\cite{ptp110549}, there are ten chromo-magnetic
penguin diagrams contributing to the $B$ decays, but only two of
them are important, as illustrated by Fig.
\ref{fig:nlodiagram}(i)and \ref{fig:nlodiagram}(j), while the other
eight diagrams are negligible.
The corresponding weak effective Hamiltonian contains the
$b\rightarrow dg$ transition:
\begin{eqnarray}\label{eq:mgheff}
\mathcal {H}^{(mg)}_{eff}=-\frac{G_F}{\sqrt{2}}\xi^*_tC^{eff}_{8g}O_{8g},
\end{eqnarray}
with
\begin{eqnarray}\label{eq:o8g}
O_{8g}=\frac{g}{8\pi^2}m_b\bar{d}_i\sigma_{\mu\nu}(1+\gamma_5)T^a_{ij}G^{a\mu\nu}b_j
\end{eqnarray}
where $i, j$ being the color indices of quarks. The corresponding effective Wilson coefficient
$C^{eff}_{8g}=C_{8g}+C_5$\cite{prd114005}.

The decay amplitudes of Fig.\ref{fig:nlodiagram}(i) and
\ref{fig:nlodiagram}(j) can be written as
\begin{eqnarray}\label{eq:mgrho}
\mathcal
{M}^{(mg)}_{V\pi}&=&\frac{4}{\sqrt{3}}C_F^2m_B^6\int_0^1dx_1dx_2dx_3\int_0^{\infty}
b_1db_1b_2db_2b_3db_3\phi_B(x_1,b_1)\nonumber\\&&\times
\left\{[(1-x_2)\phi^A_{\pi}(x_3)[2\phi_V(x_2)+r_V(3\phi^s_V(x_2)+\phi^t_V(x_2))\right.
\nonumber\\&& \left.
+x_2r_V(\phi^s_V(x_2)-\phi^t_V(x_2))]-r_{\pi}x_3(1+x_2)(3\phi^P_{\pi}(x_3)-\phi^T_{\pi}(x_3))\phi_V(x_2)\right.
\nonumber\\&& \left.
-r_{\pi}r_{V}(1-x_2)(\phi^s_V(x_2)-\phi^t_V(x_2))(3\phi^P_{\pi}(x_3)+\phi^T_{\pi}(x_3))\right.
\nonumber\\&& \left.
+r_{\pi}r_{V}x_3(1-2x_2)(\phi^s_V(x_2)+\phi^t_V(x_2))(\phi^T_{\pi}(x_3)-3\phi^P_{\pi}(x_3))]\right.
\nonumber\\&& \left. \times
C^{eff}_{8g}\alpha_s^2(t_1)h_{mg}(A,B,C,b_1,b_2,b_3)S_t(x_2)\exp[-S_{mg}]\right.
\nonumber\\&& \left.
+2r_V[2\phi^A_{\pi}(x_3)+x_3r_{\pi}(\phi^T_{\pi}(x_3)-3\phi^P_{\pi}(x_3)]\phi^s_V(x_2)\right.
\nonumber\\&& \left. \times
C^{eff}_{8g}\alpha_s^2(t_2)h_{mg}(A',B',C',b_2,b_1,b_3)S_t(x_1)\exp[-S_{mg}]
\right\},
\end{eqnarray}
\begin{eqnarray}\label{eq:mgpi}
\mathcal {M}^{(mg)}_{\pi
V}&=&\frac{4}{\sqrt{3}}C_F^2m_B^6\int_0^1dx_1dx_2dx_3\int_0^{\infty}
b_1db_1b_2db_2b_3db_3\phi_B(x_1,b_1)\nonumber\\&&\times
\left\{[(1-x_2)\phi_{V}(x_3)[2\phi^A_{\pi}(x_2)+r_{\pi}(3\phi^P_{\pi}(x_2)+\phi^T_{\pi}(x_2))\right.
\nonumber\\&& \left.
+x_2r_{\pi}(\phi^P_{\pi}(x_2)-\phi^T_V(x_2))]+r_{V}x_3(1+x_2)(3\phi^s_{V}(x_3)-\phi^t_{V}(x_3))\phi^A_{\pi}(x_2)\right.
\nonumber\\&& \left.
+r_{\pi}r_{V}(1-x_2)(3\phi^s_V(x_3)+\phi^t_V(x_3))(\phi^P_{\pi}(x_2)-\phi^T_{\pi}(x_2))\right.
\nonumber\\&& \left.
+r_{\pi}r_{V}x_3(1-2x_2)(3\phi^s_V(x_3)-\phi^t_V(x_3))(\phi^T_{\pi}(x_2)+\phi^P_{\pi}(x_2))]\right.
\nonumber\\&& \left. \times
C^{eff}_{8g}\alpha_s^2(t_1)h_{mg}(A,B,C,b_1,b_2,b_3)S_t(x_2)\exp[-S_{mg}]\right.
\nonumber\\&& \left.
+2r_{\pi}[2\phi_{V}(x_3)-x_3r_{V}(\phi^t_{V}(x_3)-3\phi^s_{V}(x_3)]\phi^P_{\pi}(x_2)\right.
\nonumber\\&& \left. \times
C^{eff}_{8g}\alpha_s^2(t_2)h_{mg}(A',B',C',b_2,b_1,b_3)S_t(x_1)\exp[-S_{mg}]
\right\},
\end{eqnarray}
where
\begin{eqnarray}
A&=&\sqrt{x_2}m_b,\quad B=B'=\sqrt{x_1x_2}m_B,\quad C=\sqrt{x_3(1-x_2)}m_B,\nonumber\\
A'&=&\sqrt{x_1}m_b,\quad C'=\sqrt{|x_1-x_3|}m_B.
\end{eqnarray}
The hard scales $t_1,t_2$ are the same as in Eq.(\ref{eq:t1t2}). The
hard function $h_{mg}$ and the Sudakov exponent $S_{mg}$ are given in the
appendix. The jet function $S_t(x_i)$ can be found in
Ref.\cite{0105003}.

\section{Numerical results and discussions}\label{sec:numerical}

 Besides those specified in the text, the following input parameters
will also be used in the numerical calculations\cite{pdg2010}:
\begin{eqnarray}\label{eq:constant}
m_B&=&5.28\text{GeV},\quad \tau_{B^0}=1.53\text{ps},\quad \tau_{B^+}=1.638\text{ps},\quad
 \nonumber\\
f_B&=&0.21\pm 0.01\text{GeV} , \quad   |V_{ub}|=(3.47^{+0.16}_{-0.12})\times 10^{-3},\quad
|V_{ud}|=0.97428,\nonumber\\
 |V_{tb}|&=&0.999,\quad
 |V_{td}|=(8.62^{+0.26}_{-0.20})\times 10^{-3},\quad \alpha=(90\pm 5)^{\circ}.
\end{eqnarray}
The corresponding values of $\Lambda_{\text{QCD}}$ are derived from
$\alpha_s(m_Z)=0.1184$ using LO and NLO formulas, respectively:
\begin{eqnarray}\label{eq:lamdaqcd}
\text{LO}: \quad\Lambda_{\text{QCD}}^{(5)}&=&(0.110 \pm
0.005)\text{GeV} ,\quad
\Lambda_{\text{QCD}}^{(4)}=0.148\text{GeV}; \nonumber\\
\text{NLO}:\quad\Lambda_{\text{QCD}}^{(5)}&=&(0.228 \pm
0.008)\text{GeV},\quad \Lambda_{\text{QCD}}^{(4)}=0.325\text{GeV}.
\end{eqnarray}

The $B$ meson distribution amplitude is given by
\begin{eqnarray}\label{eq:bwave2}
\phi_B(x,b)=N_Bx^2(1-x)^2\exp[-\frac{M_B^2x^2}{2\omega_b^2}-\frac{1}{2}(\omega_bb)^2],
\end{eqnarray}
where the shape parameter $\omega_b=0.40\pm0.04$GeV has been fixed
using the rich experimental data on the $B^0_d$ and $B^{\pm}$
decays\cite{prd63074009,prd014019,prd074018}.

For the $\pi$ meson, the twist-2 distribution amplitude
$\phi^A(x)$, and the twist-3 distribution amplitudes $\phi^P(x)$ and
$\phi^T(x)$ are written as \cite{epjc23275}
\begin{eqnarray}\label{eq:piwave1}
\phi^A_{\pi}(x)&=&\frac{3f_{\pi}}{\sqrt{2N_c}}x(1-x)[1+a_2^{\pi}C^{3/2}_2(2x-1)+0.25C^{3/2}_4(2x-1)],\nonumber\\
\phi^P_{\pi}(x)&=&\frac{f_{\pi}}{2\sqrt{2N_c}}[1+0.43C^{1/2}_2(2x-1)+0.09C^{1/2}_4(2x-1)],\nonumber\\
\phi^T_{\pi}(x)&=&\frac{f_{\pi}}{2\sqrt{2N_c}}(1-2x)[1+0.55(10x^2-10x+1)]
\end{eqnarray}
with the pion decay constant $f_{\pi}=0.13\text{GeV}$. The Gegenbauer
polynomials are defined by
\begin{eqnarray}\label{eq:Gegenbauer}
C^{1/2}_2(t)&=&\frac{1}{2}(3t^2-1),\quad
C^{1/2}_4(t)=\frac{1}{8}(35t^4-30t^2+3),\nonumber\\
C^{3/2}_2(t)&=&\frac{3}{2}(5t^2-1),\quad
C^{3/2}_4(t)=\frac{15}{8}(21t^4-14t^2+1)
\end{eqnarray}
whose coefficients correspond to $m^{\pi}_0=1.4\text{GeV}$.

The distribution amplitudes for the vector meson are listed below \cite{epjc23275}:
\begin{eqnarray}\label{eq:vwave2}
 \phi_V(x)&=&\frac{3}{\sqrt{6}}f_Vx(1-x)[1+a_{V}^{\parallel}C^{3/2}_2(2x-1)],\nonumber\\
 \phi_V^t(x)&=&\frac{f^T_V}{2\sqrt{6}}[3(2x-1)^2+0.3(2x-1)^2(5(2x-1)^2-3)
 \nonumber\\&&+0.21(3-30(2x-1)^2+35(2x-1)^4)],\nonumber\\
 \phi_V^s(x)&=&\frac{3}{2\sqrt{6}}f^T_V(1-2x)[1+0.76(10x^2-10x+1)],
\end{eqnarray}
with the decay constant $f_{\rho}=0.216\text{GeV}$,
$f^T_{\rho}=0.165\text{GeV}$,  $f_{\omega}=0.195\text{GeV}$ and
$f^T_{\omega}=0.145\text{GeV}$\cite{prd094020}.

\subsection{Branching Ratios}

The considered
 NLO contributions can interfere with the LO part constructively or destructively
 for different decay modes.
In Table \ref{tab:brvalue}, we show our pQCD results for the
CP-averaged branching ratios of the seven $B\rightarrow \pi \rho,
\pi \omega$ decays together with the experimental data.
In order to show the effects of the improvement, we use the same updated
input paraments for the LO and NLO calculations, which make
the LO-pQCD  predictions  larger
than the  previous pQCD calculations \cite{epjc23275}. Apparently, most of the NLO-pQCD
predictions agree  with the experimental measured values  and better than the LO results.

For comparison, we also list theoretical predictions based on the traditional
QCD factorization approach (QCDF-I) \cite{npb675333}, modified QCD factorization approach (QCDF-II) \cite{09095229}
which  include the fitted penguin annihilation topology and  color-suppressed tree amplitudes,   and the ones obtained
using SCET \cite{0801}. Comparing with the experimental
data \cite{pdg2010}, it is easy to see that the LO-pQCD predictions
are worse than the  QCDF results, but our NLO-pQCD results
have a better agreement with the experimental data. Our NLO
predictions of the branching ratios for $B\rightarrow
\pi^{\pm}\rho^{\mp}$ decays are close to QCDF-II result but larger than those
in SCET.  Neglecting the small terms, it is due to  the
different $B\rightarrow \pi$ and $B\rightarrow \rho$ form factors:
SCET uses the smaller form factors $F^{B\rightarrow \pi} = 0.198$
and $A_0^{B\rightarrow \rho} = 0.291 $; while in our NLO calculations,
$F^{B\rightarrow \pi} = 0.23$ and $A_0^{B\rightarrow \rho} = 0.30 $.

\begin{table}
\caption{Branching ratios $(\times10^{-6})$ of $B\rightarrow \pi
\rho, \pi \omega$ decays in the pQCD approach, together with results
from the QCDF-I \cite{npb675333}, QCDF-II\cite{09095229}, the ones
obtained from one solution of SCET \cite{0801} and the experimental
data \cite{pdg2010}.} \label{tab:brvalue}
\centering
\resizebox{16cm}{2.5cm}{
\begin{tabular*}{20cm}{@{\extracolsep{\fill}}l|ccccccc|cc}
\hline\hline
Mode & LO-pQCD    &NLO-pQCD &QCDF-I \cite{npb675333}  &QCDF-II \cite{09095229} & SCET \cite{0801} &Data \cite{pdg2010}\\[1ex]
\hline
  $B^0/\bar{B}^0\rightarrow \pi^{\pm}\rho^{\mp}$ &41.3  &$25.7^{+7.0+2.4+1.3+1.8}_{-5.5-1.9-2.0-1.6}$
  &$36.5^{+18.2+10.3+2.0+3.9}_{-14.7-8.6-3.5-2.9}$&$25.1^{+1.5+1.4}_{-2.2-1.8}$&$16.8^{+0.5+1.6}_{-0.5-1.5}$ &$23\pm 2.3$ \\
\hline
  $B^+\rightarrow \pi^{+}\rho^{0}$ &9.0 &$5.4^{+1.4+0.5+0.6+0.3}_{-1.1-0.3-0.5-0.0}$
  &$11.9^{+6.3+3.6+2.5+1.3}_{-5.0-3.1-1.2-1.1}$&$8.7^{+2.7+1.7}_{-1.3-1.4}$  &$7.9^{+0.2+0.8}_{-0.1-0.8}$ &$8.3\pm 1.2$ \\
  \hline
  $B^+\rightarrow \rho^{+}\pi^{0}$ &14.1 &$9.6_{-2.1-0.7-1.3-0.6}^{+2.5+0.8+0.7+0.7}$
  &$14.0^{+6.5+5.1+1.0+0.8}_{-5.5-4.3-0.6-0.7}$&$11.8^{+1.8+1.4}_{-1.1-1.4}$ &$11.4^{+0.6+1.1}_{-0.6-0.9}$&$10.9\pm 1.4$ \\
  \hline
  $B^0\rightarrow \rho^{0}\pi^{0}$ &0.15 &$0.37^{+0.09+0.02+0.03+0.08}_{-0.08-0.01-0.05-0.02}$
  &$0.4^{+0.2+0.2+0.9+0.5}_{-0.2-0.1-0.3-0.3}$&$1.3^{+1.7+1.2}_{-0.6-0.6}$ &$1.5^{+0.1+0.1}_{-0.1-0.1}$&$2.0\pm 0.5$ \\
  \hline
  $B^+\rightarrow \pi^{+}\omega$ &8.4 &$4.6^{+1.2+0.5+0.5+0.1}_{-0.9-0.4-0.4-0.1}$
  &$8.8^{+4.4+2.6+1.8+0.8}_{-3.5-2.2-0.9-0.9}$&$6.7^{+2.1+1.3}_{-1.0-1.1}$&$8.5^{+0.3+0.8}_{-0.3-0.8}$&$6.9\pm 0.5$ \\
  \hline
  $B^0\rightarrow \pi^{0}\omega$ &0.22&$0.32^{+0.06+0.01+0.04+0.04}_{-0.05-0.02-0.07-0.04 }$
  &$0.01^{+0.00+0.02+0.02+0.03}_{-0.00-0.00-0.00-0.00}$&$0.01^{+0.02+0.04}_{-0.00-0.01}$&$0.015^{+0.024+0.002}_{-0.000-0.002}$ &$<0.5$ \\
\hline\hline
\end{tabular*}}
\end{table}

For the color-suppressed tree dominant mode $B^0\rightarrow
\pi^{0}\rho^{0}$, the NLO pQCD contributions enhance its branching
ratio  by a factor 2.5, which are helpful to pin down the gap
between the pQCD calculations and the experimental data. This NLO  $\mathcal {BR}(B^0\rightarrow
\pi^{0}\rho^{0})$ is comparable with the result of QCDF-I,
but still smaller than QCDF-II and SCET results and the experimental data.
Soft corrections to $a_2$  enhance the QCDF-II predictions,
while in the SCET framework, the hard-scattering form factor
$\zeta_J$ is fitted to be relatively large and comparable with the
soft form factor $\zeta$.
In a very recent paper \cite{prd034023},  the authors show the existence of residual infrared divergences caused by
Glauber gluons in   non-factorizable emission diagrams,  which may resolve the large discrepancy  between the theoretical
predictions on  $\mathcal {BR}(B^0\rightarrow \pi^{0}\rho^{0})$ and the data.
 For another
color-suppressed tree dominant mode $B^0\rightarrow \pi^{0}\omega$,
our  pQCD prediction is comparable with the $B^0\rightarrow
\pi^{0}\rho^{0}$ mode; while both  QCDF and SCET predictions for
this mode are less than $B^0\rightarrow \pi^{0}\rho^{0}$ results.
This should be clarified by  future experiments.

 The theoretical
uncertainties of  the NLO-pQCD predictions are also shown in  Table
\ref{tab:brvalue}. The first error  comes from the B meson wave
function parameters $\omega_b = 0.40\pm0.04$ and $f_B = 0.21 \pm
0.01 \text{GeV}$; the second error arises from the uncertainties of
the CKM matrix elements $|V_{ub}|=(3.47^{+0.16}_{-0.12})\times
10^{-3}$,  $|V_{td}|=(8.62^{+0.26}_{-0.20})\times 10^{-3}$ and the
CKM angles $\alpha=(90\pm5)^{\circ}$; the third error comes from the
uncertainties of final state meson wave function parameters
$a_2^{\pi}=0.44^{+0.1}_{-0.2}$, $a_{\rho}^{\parallel}=0.18\pm 0.1$ \cite{zpc48239};
the fourth error is from the hard scale $t$ varying from $0.75t$ to
$1.25t$ and $\Lambda^{(5)}_{QCD}=0.228^{+0.008}_{-0.009}\text{GeV}$,
which characterizes the uncertainty of higher order contributions. It
is easy to see that the most important uncertainty   in our approach
comes from the B meson wave function and CKM elements $V_{ub}$. The
total theoretical error is in general around $30\%$ to $50\%$ in
size, which is smaller than the previous leading-order calculation.

\begin{figure}[!htbh]
\centering
\includegraphics[width=5.5in]{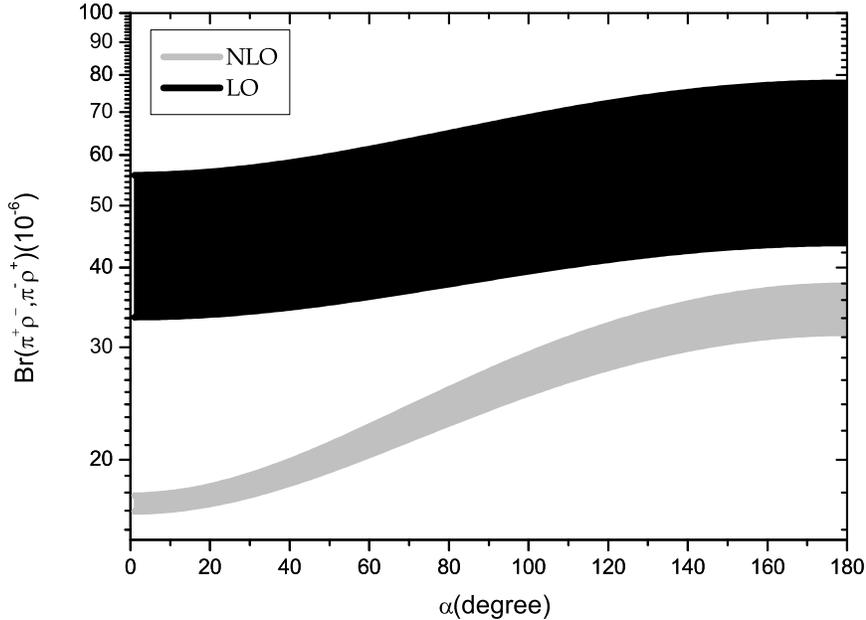}
\caption{The scale dependence of  $Br(B^{0}/\bar{B}^0\rightarrow
\pi^{\pm} \rho^{\mp})$ of the LO(the black band) and the NLO(the
gray band).}\label{fig:scale}
\end{figure}

Since both tree and penguin diagrams contribute to these decays, the
decay amplitude for a given decay mode with $\bar{b}\rightarrow
\bar{d}$ transition can be parameterized using CKM unitarity as
\begin{eqnarray}\label{eq:ampli1}
\mathcal {A}=\xi^*_uT-\xi^*_tP=\xi^*_uT[1+z e^{ i(\alpha+\delta)}],
\end{eqnarray}
where the parameter $z=|\xi_t/\xi_u||P/T|$,  the weak phase
$\alpha=\arg[-\xi_t/\xi_u]$, and $\delta=\arg[P/T]$ is the relative
strong phase between T and P part. The corresponding charge
conjugate decay mode is then
\begin{eqnarray}\label{eq:ampli2}
\mathcal {\overline{A}}=\xi_uT-\xi_tP=\xi_uT[1+z e^{
i(-\alpha+\delta)}].
\end{eqnarray}
The CP-averaged branching ratio is
\begin{eqnarray}\label{eq:branching}
\mathcal {B}r(B\rightarrow \pi \rho)=\frac{\tau_B}{16\pi
m_B}\frac{|\mathcal {A}|^2+|\mathcal {\overline{A}}|^2}{2}
=\frac{\tau_B}{16\pi m_B}|\xi_uT|^2[1+2z\cos \alpha \cos\delta+z^2],
\end{eqnarray}
which shows a clear CKM angle $\alpha$ dependence. This potentially
gives a way to measure the CKM angle $\alpha$ by these decays, if we
 can really pin down the large theoretical uncertainties of the
branching ratio calculations. For illustration, we show the LO and
NLO results of $Br(B^0/\bar{B}^0\rightarrow \pi^{\pm}\rho^{\mp})$
 in Fig \ref{fig:scale} as a function of $\alpha$ with the hard
scales varied from $0.75t$ to $1.25t$. We observe that the scale
dependence of the NLO branching ratio is significantly smaller than
that of the LO branching ratio, roughly from $\approx50\%$, reduced
to less than $10\%$.

\subsection{CP asymmetries}

\begin{figure}[!htbh]
\centering
\includegraphics[width=5.5in]{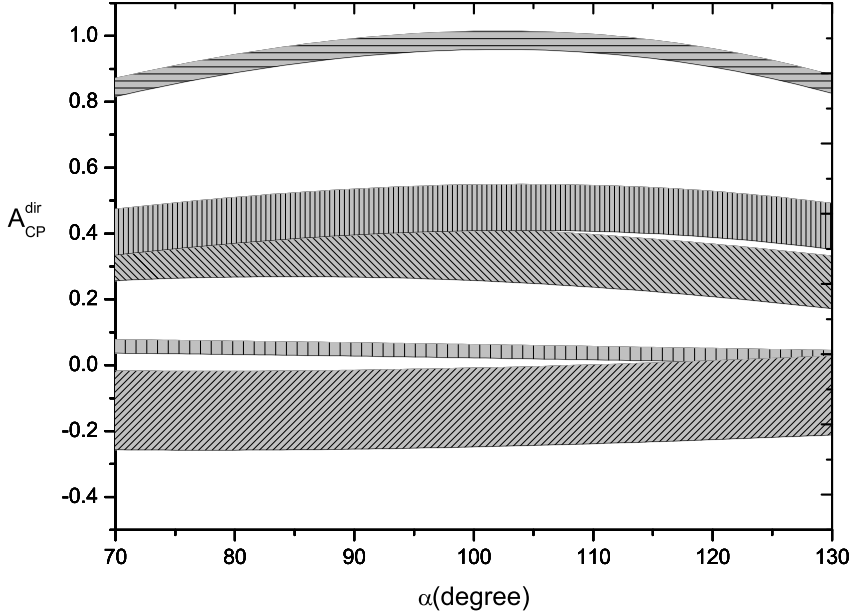}
\caption{Direct CP-violation parameters of $B^0\rightarrow \pi^0
\omega $ (the top band), $B^+\rightarrow \rho^0 \pi^0 $ (the second
band), $B^+\rightarrow \rho^+ \pi^0 $ (the third band),
$B^0\rightarrow \pi^+ \omega $ (the fourth band), $B^+\rightarrow
\pi^+ \rho^0
 $ (the bottom band), as a function of CKM angle
$\alpha$}.\label{fig:acp}
\end{figure}

Using (\ref{eq:ampli1}) and (\ref{eq:ampli2}), we can derive the
direct CP-violating parameter
\begin{eqnarray}\label{eq:acp}
A^{dir}_{CP}=\frac{\mathcal {|\overline{A}}|^2-|\mathcal
{A}|^2}{|\mathcal {A}|^2+\mathcal {|\overline{A}}|^2}= \frac{2z\sin
\alpha \sin \delta}{1+2z\cos \alpha \cos \delta+z^2}.
\end{eqnarray}
It is clear that the non-zero direct CP asymmetry requires at least
two comparable  contributions with different strong phase and
different weak phase. Since $A^{dir}_{CP}$ is proportional to $\sin
\alpha$, it can be used to measure the CKM angle $\alpha$, if we
know the strong phase difference between the tree and penguin
diagrams. The CKM angle $\alpha $ dependence of the direct
CP-violating asymmetries of these decays are shown in Fig.
\ref{fig:acp}. The accuracy of this measurement requires more
precise theoretical calculation and more experimental data.

 The numerical results for the direct CP-violating
asymmetries of $B^{\pm}\rightarrow\pi^{\pm}\rho^0$,
$\rho^{\pm}\pi^0$, $\pi^{\pm} \omega$ and
$B^{0}\rightarrow\pi^{0}\rho^0$, $\pi^{0} \omega$ decays are listed
in Table \ref{tab:acpdir}. The direct CP-violation parameters of
$B^+\rightarrow \pi^+\rho^0$ is negative,
 while the direct CP-violation parameter of the other  modes are positive. The direct
CP-violation parameter of $B^+\rightarrow \pi^+\omega$ is rather
small for the almost canceled  contributions of  annihilation diagram, which
are the dominant contributions to the strong phases in pQCD
approach. Because the NLO Wilson evolution increases the penguin
amplitudes and dilutes the tree amplitudes, the NLO direct
CP-violation parameters (absolute value)
 of those decays are slightly enhanced
 compared  with the LO predictions.
However, for the color-suppressed tree dominant modes
$B^0\rightarrow \pi^{0}\rho^{0}$ and $B^0\rightarrow \pi^{0}\omega$,
the direct CP asymmetry varies from $-50\%$ to $47\%$ and from
$52\%$ to $98\%$, respectively. The big changes   are attributed to
a huge change of the strong phase of color-suppressed tree
amplitudes caused by the vertex corrections.

\begin{table}
\caption{The pQCD predictions for the direct CP-violating
asymmetries of $B^{\pm}\rightarrow\pi^{\pm}\rho^0,\rho^{\pm}\pi^0,
\pi^{\pm} \omega$ and $B^{0}\rightarrow\pi^{0}\rho^0, \pi^{0}
\omega$ decays$ (\text{in units of }\%)$.
 We cite theoretical
results evaluated
 in QCDF-I \cite{npb675333}, QCDF-II\cite{09095229}, SCET \cite{0801} and experimental data \cite{pdg2010}
for comparison.}\label{tab:acpdir}
\resizebox{16cm}{2.5cm}{
\begin{tabular*}{20cm}{@{\extracolsep{\fill}}l|cc|cccc}
\hline\hline
Mode & LO    &NLO &QCDF-I \cite{npb675333}&QCDF-II \cite{09095229}&SCET \cite{0801} &Data \cite{pdg2010}\\[1ex]
\hline
  $B^{\pm}\rightarrow\pi^{\pm}\rho^0$ &-26.4 &$-13.2^{+4.8+0.7+6.5+8.5}_{-5.3-0.7-5.5-9.6}$
  &$4.1^{+1.3+2.2+0.6+19.0}_{-0.9-2.0-0.7-18.8}$&$-9.8^{+3.4+11.4}_{-2.6-10.2}$&$-19.2^{+15.5+1.7}_{-13.4-1.9}$ &$ 18^{+9}_{-17}$ \\
\hline
  $B^{\pm}\rightarrow\rho^{\pm}\pi^0$ &20.1&$34.7^{+4.4+1.6+4.4+8.8}_{-4.1-1.6-4.8-8.2}$
  &$-4.0^{+1.2+1.8+0.4+17.5}_{-1.2-2.2-0.4-17.7}$&$9.7^{+2.1+8.0}_{-3.1-10.3}$ &$12.3^{+9.4+0.9}_{-10.0-1.1}$ &$2\pm 11$ \\
  \hline
  $B^{\pm}\rightarrow\pi^{\pm}\omega$ &0.4&$5.3^{+0.3+0.3+1.2+0.8}_{-0.1-0.3-0.5-2.5}$
  &$-1.8^{+0.5+2.7+0.8+2.1}_{-0.5-3.3-0.7-2.2}$&$-13.2^{+3.2+12.0}_{-2.1-10.7}$ &$2.3^{+13.4+0.2}_{-13.2-0.4}$&$-4\pm 6$ \\
   \hline
  $B^{0}\rightarrow\pi^{0}\rho^{0}$ &-49.8 &$46.5_{-8.2-2.1-2.6-7.0}^{+8.4+2.2+6.2+7.1}$
  &$-15.7^{+4.8+12.3+11.0+19.8}_{-4.7-14.0-12.9-25.8}$&$11.0^{+5.0+23.5}_{-5.7-28.8}$ &$-3.5^{+21.4+0.3}_{-20.3-0.3}$&$-30\pm 40$ \\
  \hline
  $B^{0}\rightarrow\pi^{0}\omega$ &51.9&$97.6_{-1.1-2.1-1.3-3.0}^{+0.0+1.5+0.7+3.0}$
  &--&$-17.0^{+55.4+98.6}_{-22.8-82.3}$ &$39.5^{+79.1+3.4}_{-185.5-3.1}$&-- \\
\hline\hline
\end{tabular*}}
\end{table}


The theoretical uncertainties of  the NLO-pQCD predictions are also
shown in  Table \ref{tab:acpdir}. The first error shown in the
table, comes from the B meson wave function parameters $\omega_b =
0.40\pm0.04$ and $f_B = 0.21 \pm 0.01 \text{GeV}$; The second error
arises from the uncertainties of the CKM matrix elements
$|V_{ub}|=(3.47^{+0.16}_{-0.12})\times 10^{-3}$,
$|V_{td}|=(9.62^{+0.26}_{-0.2})\times 10^{-3}$ and the CKM angles
$\alpha=(90\pm5)^{\circ}$; the third error comes from the
uncertainties of final state meson wave function parameters
$a_2^{\pi}=0.44^{+0.1}_{-0.2}$, $a_{\rho}^{\parallel}=0.18\pm 0.1$;
the fourth error is from the hard scale $t$ varying from $0.75t$ to
$1.25t$ and $\Lambda^{(5)}_{QCD}=0.228^{+0.008}_{-0.009}\text{GeV}$,
characterizing the uncertainty of higher order contributions.
 Unlike the CP-averaged branching ratios, the direct CP
asymmetry is not sensitive to the wave function parameters and CKM
factors, since these parameter dependence canceled out in
Eq.(\ref{eq:acp}). In addition,  the CKM angles ($\alpha$)
uncertainty is quite small ($\sim 5\%$). Therefore, the most
important uncertainties here are  the scale
 dependence, which shows the importance of the NLO
calculations.

 We also cite results evaluated in QCDF-I \cite{npb675333}, QCDF-II \cite{09095229},
 SCET\cite{0801} for comparison in Table
\ref{tab:acpdir}. Our predictions on direct CP asymmetries
 are typically larger in magnitude, most of which have the same sign with SCET approach.
In QCDF framework,
 the strong phases are either at the order of $\alpha_s$ or power
suppressed in $\Lambda_{QCD}/m_b$. So predictions in the QCDF-I approach
 on these channels are usually small in magnitude,   most
  have different signs from our pQCD results \cite{npb675333}. In fact,
 the QCDF-II results \cite{09095229} quoted in Table \ref{tab:acpdir} already
  included large strong phase coming from penguin annihilation
  contributions, so that their results agree well with our pQCD
  ones.

\begin{figure}[!htbh]
\centering
{\includegraphics[width=5.5in]{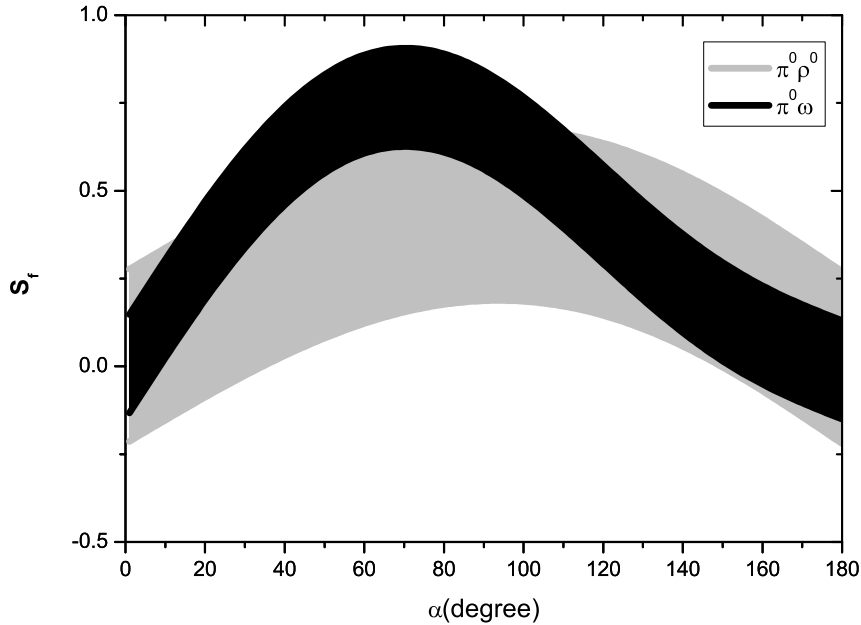}} \caption{Mixing-induced
CP-violation parameters $S_{\pi^0\rho^0}$ ( the gray band), mixing
CP-violation parameters $S_{\pi^0\omega}$ (the black band), as a
function of CKM angle $\alpha$.}\label{fig:sfacp}
\end{figure}

For the neutral $B^0$ decays, the situation is more complicated
due to the $B^0 --\bar{B}^0$ mixing. The CP asymmetry is time
dependent\cite{pdg2010}, when the final states are CP-eigenstates. A
time dependent asymmetry can be defined by
\begin{eqnarray}
A_f(t)&=&\frac{\Gamma(\bar{B}^0(t)\rightarrow
 f)-\Gamma(B^0(t)\rightarrow f)}{\Gamma(\bar{B}^0(t)\rightarrow f)+\Gamma(B^0(t)\rightarrow
f)}\\
&=&S_f\sin\Delta m t +A_{CP}^{dir} \cos \Delta m t , \label{mixcp}
\end{eqnarray}
where $\Delta m$ is the mass difference of the two mass eigenstates
 of the neutral B meson. The mixing-induced CP-asymmetry parameter $S_f$ is referred to
as
 \begin{eqnarray}
 S_f &=& \frac{2Im(\lambda_f)}{1+|\lambda_f|^2},\nonumber\\
\lambda_f&=&\frac{\xi_t}{\xi^*_t}\frac{\mathcal
{\overline{A}}}{\mathcal {A}}=
e^{2i\alpha}\frac{1+ze^{i(\delta-\alpha)}}{1+ze^{i(\delta+\alpha)}}.
\end{eqnarray}
If penguin contribution is suppressed comparing with the tree
contribution, we will have the approximate relation
 $S_f\simeq \sin2\alpha$ for a negligible
$z$ parameter. From Fig \ref{fig:sfacp}, one can see that the
 $S_f$  is not a simple $\sin2\alpha$ behavior, since the $z\simeq 3.5$
for $\pi^0\rho^0$ and $z\simeq 1.0$ for $\pi^0\omega$, reflecting a
very large  penguin contribution.

\begin{table}
\caption{The pQCD predictions for the  CP-violating parameters $S_f$
 of $B^0\rightarrow
\pi^0\rho^0,\pi^0\omega$ ({in unit of }\%), together with results
from the QCDF-II \cite{09095229}, the ones
obtained using SCET \cite{0801} and the experimental
data \cite{pdg2010}.  The errors for these entries correspond to the
uncertainties in  the scale dependence and other input parameters,
respectively.}\label{tab:acpmix}.
\begin{tabular*}{15cm}{@{\extracolsep{\fill}}l|cc|ccc}
\hline\hline
Mode & LO   &NLO  &QCDF-II \cite{09095229}&SCET \cite{0801}&Data \cite{pdg2010}\\[1ex]
\hline
  $S_{\pi^0\rho^0}$ &47  &$24^{+26+9}_{-19-12}$ &$-24^{+15+20}_{-14-22}$ &$-19^{+14+10}_{-14-15}$&$ 10\pm 40$ \\
  \hline
  $S_{\pi^0\omega}$ &-37  &$21_{-10-11}^{+5+13}$ &$78^{+14+20}_{-20-139}$&$72^{+36+7}_{-154-11}$ &-- \\
\hline\hline
\end{tabular*}
\end{table}

 The pQCD numerical results for the  CP-violating parameters
 $S_f$  of $B^0\rightarrow \pi^0\rho^0,\pi^0\omega$ are displayed in Table \ref{tab:acpmix},
together with the   QCDF-II \cite{09095229} and SCET \cite{0801}
results.
It can be seen that the pQCD central value for
    $S_{\pi^0\rho^0}$ has a different sign from the other two
    approaches, because of the penguin contribution is bigger than
    the tree contribution in our approach.
Our theoretical errors for these entries shown in the table
correspond to the uncertainties in the scale dependence and other
input parameters, respectively. It is easy to see that the
uncertainty is very large. Currently, no relevant experimental
measurements for the CP-violating asymmetries of these decays are
available. Our predictions for these quantities are different from
those in QCDF-II and SCET. We have to wait for the experimental data to
resolve these disagreements.

\subsection{Time dependent asymmetry parameters of $B^0(\bar B^0)\to
\pi^\pm\rho^\mp$ decays}

\begin{table}
\caption{The LO-and NLO-pQCD predictions for the  CP-violating
parameters $C$, $S$, $\Delta C$ and $\Delta S$
 of $B^0/\bar{B}^0\rightarrow \pi^{\pm}\rho^{\mp}$ $(\text{in units of }\%)$,   together with results
from the QCDF-I \cite{npb675333}, QCDF-II\cite{09095229}, the ones
obtained using SCET \cite{0801} and the experimental
data \cite{pdg2010}. The errors for these entries correspond to the
uncertainties in  the scale dependence and other input parameters,
respectively. }\label{tab:pirho}
\begin{tabular*}{15cm}{@{\extracolsep{\fill}}l|cc|ccccc}
\hline\hline
Mode & LO  &NLO &QCDF-I \cite{npb675333} &QCDF-II \cite{09095229}&SCET \cite{0801}&Data \cite{pdg2010}\\[1ex]
\hline
  $A_{CP}$ & -11 & $-17^{+4+4}_{-3-4}$ & $1^{+0+1+0+10}_{-0-1-0-10}$&$-11^{+0+7}_{-0-5}$&$-21^{+3+2}_{-2-3}$ &$-13\pm 4$ \\
\hline
  $C$ &6 &$15^{+2+2}_{-2-2}$& $0^{+0+1+0+2}_{-0-1-0-2}$&$9^{+0+5}_{-0-7}$ &$1^{+9+0}_{-10-0}$ &$ 1\pm 14$ \\
 \hline
  $S$ &-12  &$-31^{+6+16}_{-3-15}$& $13^{+60+4+2+2}_{-65-3-1-1}$&$-4^{+1+10}_{-1-9}$ &$-1^{+6+8}_{-7-14}$&$ 1\pm 9$\\
 \hline
  $\Delta C$ &17  &$26_{-2-8}^{+2+5}$& $16^{+6+23+1+1}_{-7-26-2-2}$&$26^{+2+2}_{-2-2}$ &$12^{+9+1}_{-10-1}$&$ 37\pm 8$\\
    \hline
  $\Delta S$ &-7  &$-7^{+0+2}_{-0-1}$& $-2^{+1+0+0+1}_{-0-1-0-1}$&$-2^{+0+3}_{-0-2}$ &$43^{+5+3}_{-7-3}$&$ -5\pm 10$\\
\hline\hline
\end{tabular*}
\end{table}

Both $B^0$ and $\bar{B}^0$ can decay into both the $\pi^{+}\rho^{-}$
and $\pi^{-}\rho^{+}$ final states. This is an interesting example
of CP asymmetry in B decays, which is the only measured combination
of four channels. $A_f$, $\bar{A}_f$, $A_{\bar{f}}$ and $\bar{A}_{\bar{f}}$ are defined as follows
\cite{npb361141}:
\begin{eqnarray}\label{eq:af}
A_f&=&\langle  \pi^-\rho^+|H_{eff}|B^0 \rangle,\quad A_{\bar{f}}=\langle \pi^+\rho^-|H_{eff}|B^0\rangle,\quad \nonumber\\
\bar{A}_f&=&\langle  \pi^-\rho^+|H_{eff}|\bar{B}^0 \rangle,\quad
\bar{A}_{\bar{f}}=\langle \pi^+\rho^-|H_{eff}|\bar{B}^0\rangle .
\end{eqnarray}
The system of four decay modes can define the time- and
flavor-integrated charge asymmetry:
\begin{eqnarray}
A_{CP}=\frac{|A_f|^2+|\bar{A}_f|^2-|A_{\bar{f}}|^2-|\bar{A}_{\bar{f}}|^2}{|A_f|^2+|
\bar{A}_f|^2+|A_{\bar{f}}|^2+|\bar{A}_{\bar{f}}|^2}.
\end{eqnarray}
 In the standard
approximation, which neglects CP violation in the $B^0-\bar{B}^0$
mixing matrix and the width difference of the two mass eigenstates,
the four time dependent widths are given by the following
formulas \cite{epjc23275}:
\begin{eqnarray}\label{eq:shuai}
\Gamma(B^0(t)\rightarrow \pi^-\rho^+)&=&e^{-\Gamma t}\frac{1}{2}(|A_f|^2+|\bar{A}_f|^2)[1+C_f\cos\Delta m t-S_f\sin\Delta m t],
\nonumber\\
\Gamma(\bar{B}^0(t)\rightarrow \pi^+\rho^-)&=&e^{-\Gamma t}\frac{1}{2}(|A_{\bar{f}}|^2+|\bar{A}_{\bar{f}}|^2)
[1-C_{\bar{f}}\cos\Delta m t+S_{\bar{f}}\sin\Delta m t],\nonumber\\
\Gamma(B^0(t)\rightarrow \pi^+\rho^-)&=&e^{-\Gamma t}\frac{1}{2}(|A_{\bar{f}}|^2+|\bar{A}_{\bar{f}}|^2)
[1+C_{\bar{f}}\cos\Delta m t-S_{\bar{f}}\sin\Delta m t],\nonumber\\
\Gamma(\bar{B}^0(t)\rightarrow \pi^-\rho^+)&=&e^{-\Gamma t}\frac{1}{2}(|A_f|^2+|\bar{A}_f|^2)[1-C_f\cos\Delta m t+S_f
\sin\Delta m t],
\end{eqnarray}
where  $\Delta m>0$ denotes the mass difference, and $\Gamma$ is the common total width of the B meson eigenstates. $C_f$
and $S_f$
are defined  as
\begin{eqnarray}\label{eq:cfsf1}
C_f&=&\frac{|A_f|^2-|\bar{A}_f|^2}{|\bar{A}_f|^2+|A_f|^2}, \quad S_f=\frac{2Im(\lambda_f)}{1+|\bar{A}_f/A_f|^2},\quad
\lambda_f=\frac{\xi_t}{\xi^*_t}\frac{\bar{A}_f}{A_f},
\end{eqnarray}
 For
decays to the CP-conjugate final state, one replaces $f$ by
$\bar{f}$ to obtain the formula for $C_{\bar f}$ and $S_{\bar f}$.
Furthermore, we define $C\equiv \frac{1}{ 2}(C_f+C_{\bar{f}})$,
$S\equiv \frac{1}{2}(S_f+S_{\bar{f}})$,
 $\Delta C\equiv \frac{1}{ 2}(C_f-C_{\bar{f}})$
and $\Delta S\equiv \frac{1}{2}(S_f-S_{\bar{f}})$. S is referred to
as mixing-induced CP asymmetry and C is the direct CP asymmetry,
while
$\Delta C$ and $\Delta S$ are CP-even under CP transformation
$\lambda_f\rightarrow 1/\lambda_{\bar{f}}$. If $f$ is CP eigenstate
there are only two different amplitudes since $f=\bar{f}$, and
$\Delta C$, $\Delta S$ vanish. The complicated formulas
(\ref{eq:shuai}) return back to the simpler one in Eq.(\ref{mixcp}).



According to (\ref{eq:ampli1}) and (\ref{eq:ampli2}), we can write
Eq.(\ref{eq:af}) as
\begin{eqnarray}\label{eq:cfsf11}
A_f&=&\xi_uT-\xi_tP,\quad A_{\bar{f}}=\xi_uT'-\xi_tP',\nonumber\\
\bar{A}_{f}&=&\xi^*_uT'-\xi^*_tP',\quad
\bar{A}_{\bar{f}}=\xi^*_uT-\xi^*_tP ,
\end{eqnarray}
where T and P denote the tree diagram amplitude and penguin diagram
amplitude of $B^0\rightarrow \rho^+\pi^-$, respectively; while
$\text{T}'$ and $\text{P}^{\prime}$ denote the tree diagram
amplitude and penguin diagram amplitude of $B^0\rightarrow
\pi^+\rho^-$, respectively. The asymmetries $\Delta S\approx
\frac{2|T||T'|}{|T|^2+|T'|^2}\sin \theta \cos \alpha$ are suppressed
by the small penguin-to-tree ratios
($|\text{P}/\text{T}|,|\text{P}'/\text{T}'|\ll 1$) and the small
relative phase $\theta$ between $\text{T}$ and $\text{T}'$ ($\theta
\simeq 3.4^{\circ}$), hence they are always small in pQCD
factorization. This conclusion is similar to that in
QCDF \cite{npb675333,09095229}, although the absolute magnitude of $\Delta S$
 are  much larger in pQCD than in QCDF. All the
CP-violation parameters of $B^0/\bar{B}^0\rightarrow
\pi^{\pm}\rho^{\mp}$ decays including the LO \cite{epjc23275} and
NLO results of pQCD,  QCDF-I \cite{npb675333}, QCDF-II \cite{09095229}, SCET \cite{0801} and the
experimental data are collected in Table \ref{tab:pirho}.
 It is clear that the NLO-pQCD
 prediction for the CP-violation parameter $A_{CP}$, $\Delta C$ and $\Delta S$ agrees with the experimental results  very well.
 The predictions of pQCD  for CP-violation parameters
 in Table \ref{tab:pirho} are comparable with the QCDF-II,
  and are better than   QCDF-I  and SCET predictions,
 which is also shown
in other B decay channels \cite{direct}.

\section{ conclusion}

In the framework of the pQCD approach, we calculated the NLO QCD
corrections to the $B\rightarrow \pi\rho$, $\pi\omega$ decays
including the vertex corrections, the quark loops,  the magnetic
penguin, and the NLO Wilson coefficients, the Sudakov factor and RG
factor. We found that the NLO corrections improved the scale
dependence significantly, and had great effects on some of the decay
channels. Our NLO-pQCD calculations agree well with the measured
values. For example, compared with LO predictions,
 the NLO corrections decease (increase) the branching ratio of
$B^0/\bar{B}^0\rightarrow \pi^{\pm}\rho^{\mp}(B^0\rightarrow \pi^0\rho^0)$, and improve
the consistency of the pQCD predictions.
The NLO corrections play an important role in modifying direct CP
asymmetries. For the color-allowed tree dominant modes, the NLO Wilson coefficients
enhance the penguin amplitudes, the larger subdominant penguin amplitudes increase
the magnitudes of the direct CP asymmetries due to the
 stronger interference with the dominant tree amplitudes.
The predictions of pQCD  for CP-violation parameters are better than
 QCDF-I  and SCET predictions.

\begin{acknowledgments}
We thank Yu Fusheng, Hsiang-nan Li, Xin Liu and Wei Wang for helpful
discussions. This work is partially supported by National Natural
Science Foundation of China under the Grant No. 10735080, and
11075168; Natural Science Foundation of Zhejiang Province of China,
  Grant No. Y606252 and Scientific Research Fund of Zhejiang Provincial Education Department of China, Grant No. 20051357;
 and the China Postdoctoral Science Foundation under
grant No. 20100480466.
\end{acknowledgments}

\begin{appendix}

\section*{Appendix}

We show here the hard function $h_{ql}$ and $h_{mg}$ the Sudakov
exponents $S_{ql, mg}(t)$ appearing in the expressions of the decay
amplitudes in \ref{sec:nlo},
\begin{eqnarray}
h_{ql}(x_1,x_2,b_1,b_2)&=&K_0(\sqrt{x_1x_2}m_Bb_1)\nonumber\\&&\times[\theta(b_1-b_2)K_0(\sqrt{x_2}m_Bb_1)I_0(\sqrt{x_2}m_Bb_2)
\nonumber\\&&+\theta(b_2-b_1)K_0(\sqrt{x_2}m_Bb_2)I_0(\sqrt{x_2}m_Bb_1)]S_t(x_2),
\end{eqnarray}
\begin{eqnarray}
h_{mg}(A,B,C,b_1,b_2,b_3)&=&-K_0(Bb_1)K_0(Cb_3)\times
\int_0^{\pi/2}d\theta \tan\theta\nonumber\\&&
J_0(Ab_1\tan\theta)J_0(Ab_2\tan\theta)J_0(Ab_3\tan\theta)
\end{eqnarray}
where $J_0$ is the Bessel function and $K_0$, $I_0$ are modified
Bessel functions with $K_0(-ix)= -(\pi/2)Y_0(x)+i(\pi/2)
J_0(x)$.

The Sudakov exponents used in the text are defined by
\begin{eqnarray}
S_{ql}(t)=s(x_1m_B,b_1)+s(x_2m_B,b_2)+s((1-x_2)m_B,b_2)+\frac{5}{3}g_2(t,b_1)+2g_2(t,b_2),
\end{eqnarray}
\begin{eqnarray}
S_{mg}(t)&=&s(x_1m_B,b_1)+s(x_2m_B,b_2)+s((1-x_2)m_B,b_2)+s(x_3m_B,b_3)\nonumber\\&&
+s((1-x_3)m_B,b_3)+\frac{5}{3}g_2(t,b_1)+2g_2(t,b_2)+2g_2(t,b_3)
\end{eqnarray}
where the functions $s(P, b)$ have been defined in
Ref.\cite{prd527}. The RG factor $g_2(t,b)$ is given by
\begin{eqnarray}
g_2(t,b)&=&-\frac{2}{\beta_0}\ln[\frac{\ln(t/\Lambda_{QCD})}{-\ln(b\Lambda_{QCD})}]+\frac{2\beta_1}{\beta^3_0}
[\frac{\ln(\ln(\frac{1}{b^2\Lambda^2_{QCD}}))}{\ln(\frac{1}{b^2\Lambda^2_{QCD}})}-
\frac{\ln(\ln(\frac{t^2}{\Lambda^2_{QCD}}))}{\ln(\frac{t^2}{\Lambda^2_{QCD}})}\nonumber\\&&+
\frac{1}{\ln(\frac{1}{b^2\Lambda^2_{QCD}})}-\frac{1}{\ln(\frac{t^2}{\Lambda^2_{QCD}})}]\nonumber\\
\beta_0&=&11-\frac{2}{3}n_f,\quad \beta_1=102-\frac{38}{3}n_f
\end{eqnarray}
where $n_f$ is the number of quarks with mass less than the energy scale $t$.
\end{appendix}



\begin{thebibliography}{99}
\bibitem{iiba}
I.I.Bigi, A.I.Sanda, CP violation, Cambridge.

\bibitem{lhcb1}
P.~Ball {\it et al.},  CERN Yellow Report 2000-004; hep-ph/0003238.

\bibitem{lhcb2}
Mario Antonelli et al. Phys. Rept. \textbf{494} (2010) 197-414.


\bibitem{prl831914}
 M. Beneke, G. Buchalla, M. Neubert, and C.T. Sachrajda, Phys. Rev. Lett. \textbf{83}, 1914
(1999); Nucl. Phys. B \textbf{591}, 313 (2000).
\bibitem{prd63074009}
 C.D. L\"{u}, K. Ukai and
M.Z. Yang, Phys. Rev. D {\bf 63}, 074009 (2001).

\bibitem{pdg2010}
Particle Data Group,  J.Phys.G: Nucl.Part. Phys. \textbf{37}, 075021
(2010).

\bibitem{pikpuzzle}S. Baek,  C.-W. Chiang,   D. London,
 Phys. Lett. B \textbf{675}, 59-63 (2009) and references therein.


\bibitem{prd114005}
 H.N. Li, S. Mishima, A.I. Sanda, Phys. Rev. D \textbf{72}, 114005 (2005).
\bibitem{prd094020}
H.N. Li and S. Mishima, Phys. Rev. D \textbf{74}, 094020 (2006);
H.N. Li and S. Mishima, Phys. Rev. D \textbf{73}, 114014 (2006).

\bibitem{0807}
Z.Q. Zhang and Z.J. Xiao,  Eur. Phys. J. C \textbf{59}, 49 (2009); arXiv:
0807.2024 [hep-ph].
\bibitem{prd114001}
Z.J. Xiao, Z.Q. Zhang, X. Liu, and L.B. Guo, Phys. Rev. D
\textbf{78}, 114001 (2008).
\bibitem{exp}
A. Kusaka et al., Belle Collaboration, Phys. Rev. D \textbf{77}, 072001(2008);
BaBar Collaboration (G Mohanty for the collaboration),
talk at 5th International Workshop on the CKM Unitarity Triangle
(CKM 2008), Rome, Italy, 9¨C13 September 2008.
\bibitem{epjc23275}
C.D. L\"{u}, M.Z. Yang, Eur. Phys. J. C \textbf{23}, 275 (2002).
\bibitem{rmp681125}
G. Buchalla, A.J. Buras, M.E. Lautenbacher, Rev. Mod. Phys.
\textbf{68}, 1125 (1996).
\bibitem{prd63114020}
C.W.Bauer, S. Fleming, D. Pirjol and I. W. Stewart, Phys. Rev. D
\textbf{63}, 114020 (2001); C.W.Bauer, D.
Pirjol and I. W. Stewart, Phys. Rev. Lett. \textbf{87}, 201806
(2001).

\bibitem{prd69094018}
Yong-Yeon Keum, T. Kurimoto, H.-n. Li, C.D. L\"{u} and A.I. Sanda,
Phys. Rev. D \textbf{69}, 094018
(2004).
\bibitem{prd074004}
B. Melic, B. Nizic, and K. Passek, Phys. Rev. D \textbf{60}, 074004
(1999).
\bibitem{qiaocf}H. Kawamura,   J. Kodaira, C.F. Qiao,   K. Tanaka,
  Phys. Lett. B \textbf{523}, 111 (2001), Erratum-ibid. B \textbf{536}, 344 (2002); Mod. Phys. Lett. A \textbf{18}, 799
(2003).





\bibitem{epjc28515}
C.D. L\"{u}, M.Z. Yang, Eur. Phys. J. C \textbf{28}, 515 (2003).
\bibitem{zpc48239}
V.M. Braun and I.E. Filyanov , Z. Phys. C \textbf{48}, 239 (1990);
P. Ball, V.M. Braun, Y. Koike, and K. Tanaka, Nucl. Phys. B
\textbf{529}, 323 (1998); P. Ball, J. High Energy Phys. \textbf{01},
010 (1999).

\bibitem{prl43242}
M. Bander, D. Silverman and A. Soni, Phys. Rev. Lett. \textbf{43},
242 (1979); J.M. Gerard and W.S. Hou, Phys. Rev. D \textbf{43}, 2909
(1991).








\bibitem{ptp110549}
S. Mishima and A.I. Sanda, Prog. Theor. Phys. \textbf{110}, 549
(2003).
\bibitem{0105003}
T. Kurimoto, H.-n. Li, A.I. Sanda, Phys.Rev. D \textbf{65}  014007 (2002).


\bibitem{prd014019}
H.-n. Li, Phys. Rev. D \textbf{64}, 014019 (2001); Phys. Rev. D
\textbf{66}, 054013 (2002); Y.-Y. Keum and A.I. Sanda, Phys. Rev. D
\textbf{67}, 054009 (2003).
\bibitem{prd074018} A. Ali, G. Kramer, Y. Li, C.D. L\"{u}, Y.L.
Shen, W. Wang and Y.M. Wang, Phys. Rev. D \textbf{76}, 074018
(2007).
\bibitem{npb675333}
 Martin Beneke and Matthias Neubert, Nucl. Phys. B
\textbf{675}, 333 (2003).
\bibitem{09095229}
Hai-Yang Cheng and Chun-Khiang Chua, Phys.Rev. D \textbf{80}, 114008(2009).

\bibitem{0801}
Wei Wang, Yu-Ming Wang, De-Shan Yang and C.D. L\"{u}, Phys. Rev. D
\textbf{78}, 034011 (2008).

\bibitem{prd034023}
H.-n. Li and S. Mishima, Phys. Rev. D \textbf{83}, 034023 (2011).


\bibitem{npb361141}
R. Aleksan, I. Dunietz, B. Kayser, F. Le Diberder, Nucl. Phys.
\textbf{B} \textbf{361} (1991) 141.



\bibitem{direct}
B.-H. Hong,   C.-D. L\"{u}, Sci. China G \textbf{49}, 357 (2006).
\bibitem{prd527}
Hsiang-nan Li,  Phys. Rev. D \textbf{52}, 7 (1995).



\end{thebibliography}
\end{document}